# Tailoring topological transition of anisotropic polaritons by interface engineering in biaxial crystals

*Yali Zeng*[1,2†], *Qingdong Ou*[3†], *Lu Liu*[4], *Chunqi Zheng*[2], *Ziyu Wang*[5], *Youning Gong*[6], *Xiang Liang*[7], *Yupeng Zhang*[6], *Guangwei Hu*[2], *Zhilin Yang*[1], *Cheng-Wei Qiu*[2], *Qiaoliang Bao*[3*], *Huanyang Chen*[1*], *Zhigao Dai*[4*]

[1] Department of Physics, Xiamen University, Xiamen, 361005, P. R. China.

[2] Department of Electrical and Computer Engineering, National University of Singapore, Singapore, 117583, Singapore

[3] Department of Materials Science and Engineering, and ARC Centre of Excellence in Future Low-Energy Electronics Technologies (FLEET), Monash University, Clayton, Victoria, Australia

[4] Engineering Research Center of Nano-Geomaterials of Ministry of Education, Faculty of Materials Science and Chemistry, China University of Geosciences, Wuhan, 430074, P. R. China.

[5] Institute of Materials Research and Engineering Agency for Science Technology and Research (A*STAR), Singapore, 138634, Singapore

[6] Institute of Microscale Optoelectronics, Shenzhen University, Shenzhen, 518060, P. R. China.

[7] School of Energy and Power Engineering, Wuhan University of Technology, Wuhan, 430063, P. R. China.

* Correspondence: qiaoliang.bao@gmail.com; kenyon@xmu.edu.cn; daizhigao@cug.edu.cn

† Yali Zeng and Qingdong Ou contributed equally to this work.

**ABSTRACT**: Polaritons in polar biaxial crystals with extreme anisotropy offer a promising route to manipulate nanoscale light-matter interactions. The dynamical modulation of their dispersion is great significance for future integrated nano-optics but remains challenging. Here, we report a momentum-directed strategy, a coupling between the modes with extra momentum supported by the interface and in-plane hyperbolic polaritons, to tailor topological transitions of anisotropic polaritons in biaxial crystals. We experimentally demonstrate such tailored polaritons at the interface of heterostructures between graphene and α-phase molybdenum trioxide (α-$MoO_3$). The interlayer coupling can be electrically modulated by changing the Fermi level in graphene, enabling a dynamic topological transition. More interestingly, we found that the topological transition occurs at a constant Fermi level when tuning the thickness of α-$MoO_3$. The momentum-directed strategy implemented by interface engineering offers new insights for optical topological transitions, which may shed new light for programmable polaritonics, energy transfer and neuromorphic photonics.

## INTRODUCTION

Polaritons–hybrid quasiparticles with photons–provide a unique way to harness and manipulate light at nanoscale due to the strong light-matter interaction. Conventional surface plasmon polaritons (PPs) propagating along the interface of metal and dielectric are widely studied with intrinsic electronic scattering and plasmonic loss[1-7]. The emergent polaritons in van der Waals (vdW) materials feature the low loss and ultrahigh confinement, including PPs in graphene[8-10] and black phosphorus (BP)[11, 12], phonon polaritons (PhPs) in hexagonal boron nitride (hBN)[13-15] and α-phase molybdenum oxide (α-$MoO_3$)[16-19], exciton polaritons in transition metal dichalcogenides[20-22], promising the integrated and ultrathin nanophotonic devices. Those great promise should rely on their controllable propagation characteristics as dominated by the dispersion of polaritons. For instance, polaritons in biaxial crystals[23-25] have the direction-related momentum, exhibiting interesting and distinctive nano-optical phenomena, such as highly directional propagation, topological transition (TT)[26] and canalization[27]. Recently, the widespread interest has been studying the polaritons at the interface between biaxial crystals and dielectric background[28], since it offers the direct access and manipulation of the wave at the surface. Such responses can also be divided into in-plane or out-of-plane ones, depending on the distinguishability of propagation characteristics along different directions at the interface. An example is that polaritons at the top surface of exfoliated hBN or graphene are in-plane isotropic with a circular wavefront, but patterning them into artificially biaxial metasurfaces with periodic array of nanoribbons can induce the in-plane anisotropy and even a hyperbolic-to-elliptic TT[29-32]. Furthermore, one can stack naturally biaxial crystals such as α-$MoO_3$ thin flakes and manipulate the interlayer coupling via their twisted angle to stimulate such TT[33-36]. Those tailorable polaritons have represented an important step towards the on-demand control of polaritons.

Biaxial crystals heterostructures provide another promising route to engineer the dispersion of polaritons. To configure the in-plane anisotropy of PhPs in hBN, stacking of hBN onto anisotropic BP has proved to induce an in-plane ellipticity of PhP dispersions[37]. Through anisotropy-oriented mode couplings and TT, the in-plane hyperbolic response of PhPs has been controlled and switched at the heterostructural interfaces between α-$MoO_3$ and SiC, enabling the propagation of hyperbolic PhPs along originally forbidden directions[38, 39]. These examples offer further approaches to steer the dispersion of polaritons. However, all those engineering tool lacks the dynamic tunability. The active tuning of anisotropic polaritons and their TTs is highly desired for integrated photonic circuits but remains unexplored.

In this work, we show the momentum-directed polaritons and tunable TTs in biaxial crystals enabled by interface engineering. These transitions are induced by the coupling between the modes with extra momentum supported by the interface and in-plane hyperbolic polaritons excited in biaxial crystals, which can be modulated by both the stack and magnitude of the interface momentum, as well as by the thickness of biaxial crystals. With natural biaxial crystal α-$MoO_3$ and isotropic graphene, we demonstrate tunable momentum-directed polaritons at the interface of graphene and α-$MoO_3$, especially for the TT of hybrid polaritons. In this hybrid system, we theoretically predict that tailored interlayer coupling enables control of the TT of polariton dispersions, and experimentally demonstrate reconfigurable polaritons at the heterostructural interfaces by real-space nano-imaging. Due to the tunability of Fermi levels in graphene, the approach manifests as a new strategy to dynamically adjust the interlayer coupling to achieve the manipulation of optical TT. Moreover, tuning of the thickness of α-$MoO_3$ slab and thereby the hyperbolic PhPs dispersions could be another dimension for controlling TTs of anisotropic polaritons.

## RESULTS AND DISCUSSION

Due to their intrinsic anisotropy, biaxial crystals can support in-plane elliptical and hyperbolic polaritons according to the signs of the in-plane permittivity tensors. To tailor the in-plane polariton dispersions, an additional momentum $k$ can be applied at the interfaces of the biaxial crystal slab to couple with polaritons in biaxial crystals (see three situations in Figure 1a, $k$ at upper (I), lower (II), or both interfaces (III)). Dynamic regulation of polariton dispersions in the biaxial crystal can be achieved by an adjustable magnitude of the applied momentum $k$ at the interface (Figure 1b). For polar biaxial crystals, the momentum of polaritons largely depends on the thickness $d$ of the material, i.e., $k_i \approx -(\varepsilon_{sup}+\varepsilon_{sub})/(d\varepsilon_{ii})$, $i = x, y$, where $\varepsilon_{sup}$ and $\varepsilon_{sub}$ are the complex dielectric function of the environment media[16]. Therefore, we can also control polariton dispersions at the interface by varying the thickness of biaxial crystals (Figure 1c).

Our momentum-directed strategy can approximatively achieve elliptic-to-circular (Figure 1d-f) and hyperbolic-to-elliptic (Figure 1g-i) polariton dispersion transformations for in-plane elliptic and hyperbolic biaxial crystals, respectively. An additional momentum $k$ is applied at the interface by adding a conductive sheet for calculations (see in Supplementary note) in Figure 1d-i. For elliptic dispersion biaxial crystals, we define the polariton anisotropy $\alpha = k_{[100]}/k_{[001]}$[37], and the more α approach to 1, the greater the in-plane isotropy. Specifically, for the lower, upper, and both interfaces, α is calculated to be approximately 0.73, 0.81, and 0.84, respectively, which is larger than 0.66 of the pristine biaxial crystal (Figure 1d). The degree of dispersion control is different for $k_{up}$ or $k_{low}$ with the same conductivity (red and green curves in Figure 1d), due to the $SiO_2$/Si substrate with broken symmetry in the $z$-direction. We also found that the in-plane isotropy α increase with the increase of the applied conductivity (Figure 1e) and the decrease of the thickness of the biaxial crystal slab with a fixed $k_{up}$ (Figure 1f).

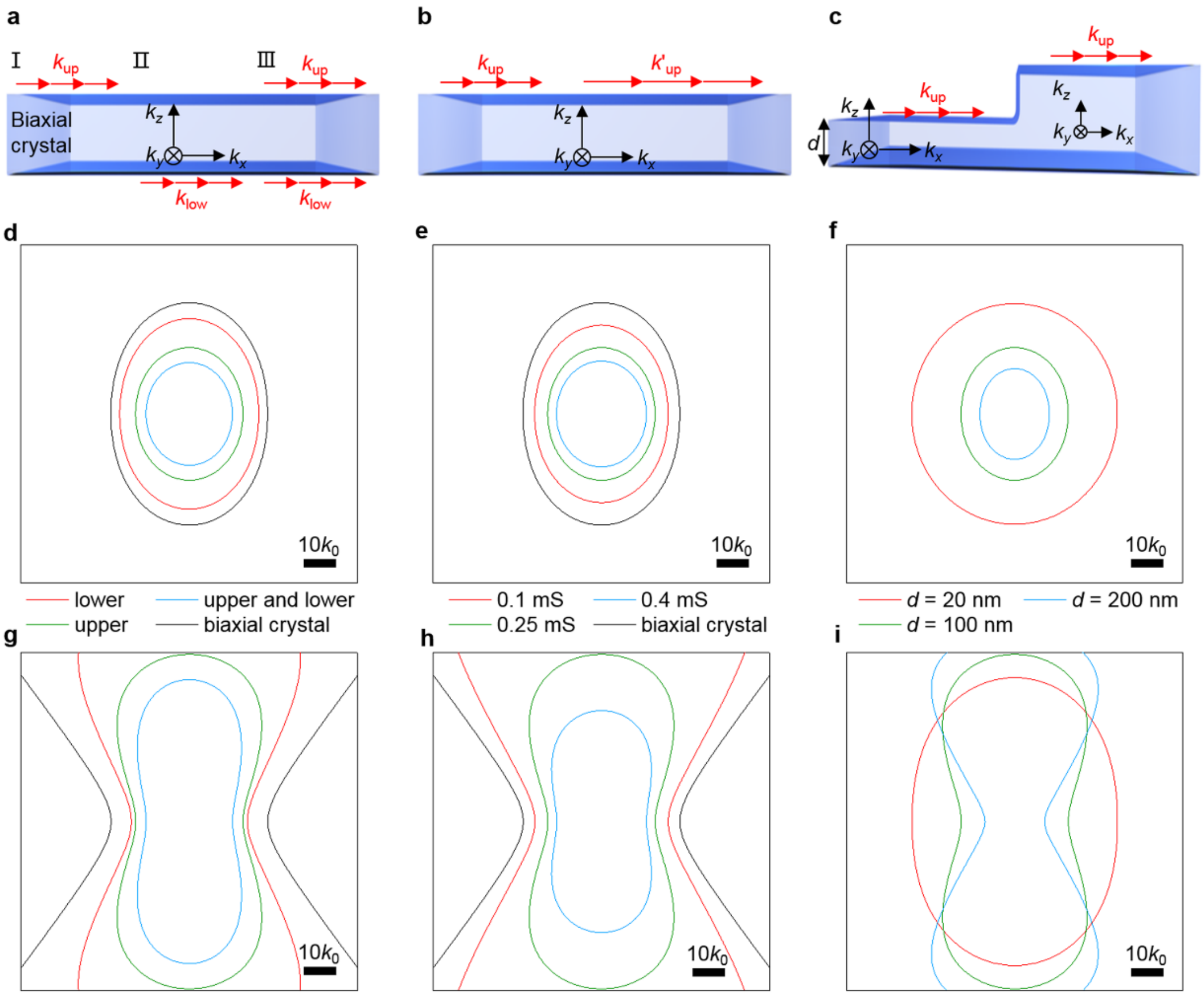

**Figure 1.** Momentum-directed polaritons in biaxial crystal. (a) Schematic of added momenta in the upper, lower, both upper and lower interface of the biaxial slab, respectively. (b) Schematic of an alterable momentum in the upper interface. (c) Schematic of a momentum in the upper interface of the biaxial slab with varied thicknesses. $k_x$, $k_y$, $k_z$ are the momenta of biaxial crystal, respectively, and $k_{up}$ and $k_{low}$ are the added momenta in upper and lower interfaces of biaxial slab, respectively. Isofrequency curves for the fundamental modes of the momentum-directed polaritons in biaxial crystal ($\varepsilon_x$ = -3, $\varepsilon_y$ = -2): (d) $d$ = 100 nm, $\sigma$ = 0.25 mS; (e) $d$ = 100 nm and added momentum in the upper interface of biaxial slab; (f) $\sigma$ = 0.25 mS and added momentum in the upper interface of biaxial slab. Isofrequency curve for the momentum-directed polaritons in biaxial crystal ($\varepsilon_x$ = -3, $\varepsilon_y$ = 2): (g) $d$ = 100 nm, $\sigma$ = 0.25 mS; (h) $d$ = 100 nm and added momentum in the upper interface of biaxial slab; (i) $\sigma$ = 0.25 mS and added momentum in the upper interface of biaxial slab.

For in-plane hyperbolic biaxial crystals, the momentum-directed hybrid polaritons behave remarkably differently. In the case of $k$ at upper interface, the isofrequency dispersion of the hybrid polariton in low $k$ changes to a closed curve (green curve in Figure 1g), while the open (hyperbolic) dispersion preserves with $k_{\text{low}}$ (red curve in Figure 1g). For both $k_{\text{up}}$ and $k_{\text{low}}$, the isofrequency curve of the hybrid polariton is an approximate ellipse. The TT from hyperbola to ellipse originates from the hybridization between the interface momentum and hyperbolic polaritons in the biaxial crystal. A similar TT arises when the conductivity increase (Figure 1h). The cutoff wavevectors of the hybrid polaritons along [100] and [001] directions increase as the conductivity increases. For thickness modulation, we can find that $k_{[100]}$ increases gradually, but $k_{[001]}$ decreases gradually as the thickness decrease, and the isofrequency curve gradually changes from concave to convex (Figure 1i). Due to the intriguing TTs observed for hyperbolic dispersions, we focus on the manipulation of polaritons in in-plane hyperbolic biaxial crystals in the following discussions.

To demonstrate the momentum-directed polaritons, we select graphene as a tunable extra momentum source due to its flexible tunability, easy of fabrication and stacking, and biaxial crystal α-$MoO_3$ to study the interlayer coupling for tailoring TT of anisotropic polaritons. To directly show the overlapping dielectric responses, the real parts of the permittivity tensors of graphene (see method) and α-$MoO_3$ along the three principal axes[17] are plotted in Figure 2a, suggesting the potential formation of hybrid plasmon-phonon polaritons modes in two excitation bands (Figure2a, 545 ~ 820 $cm^{-1}$, 851 ~ 972 $cm^{-1}$). Three configurations were proposed to fully engineer the interface of graphene/α-$MoO_3$ heterostructures (see the inserts in Figure 2b-d): α-$MoO_3$/graphene heterostructure ($MoO_3$/G), graphene/α-$MoO_3$ heterostructure (G/$MoO_3$) and graphene/α-$MoO_3$/graphene sandwich heterostructure (G/$MoO_3$/G), all with the $SiO_2$/Si substrates. The simulated field distribution Re($E_z$) for the $MoO_3$/G heterostructure is shown in Figure 2b, exhibiting hybrid plasmon-phonon polaritons with an evident concave wavefront. Here, the doping level of graphene is $E_F$ = 0.3 eV. The corresponding in-plane dispersion of the hyperbolic hybrid polaritons can be viewed more intuitively by the fast Fourier transform (FFT) of Re($E_z$), (Figure 2e), which has an excellent agreement with the analytical dispersion curve (white solid curves). By contrast, with the same Femi level, hybrid polaritons in G/$MoO_3$ heterostructure (Figure 2c) show a closed (elliptical) dispersion contour, indicating finite and direction-dependent in-plane wavevectors (Figure 2f). For the G/$MoO_3$/G sandwich heterostructure (Figure 2d), a closed dispersion contour is also obtained (Figure 2g), but with a smaller cutoff of in-plane wavevector compared to Figure 2f. The trend of the isofrequency curves of momentum-directed hybrid phonon-plasmon polaritons in Figure 2e-g agrees well with that in Figure 1g. To exclude the substrate effect, the field distributions and corresponding dispersions of suspended heterostructures are also simulated (Figure S2), indicating no difference for the suspended G/$MoO_3$ and $MoO_3$/G heterostructures. Those results are very interesting as it suggests the importance of stack when we break the symmetry of superstrate and substrate.

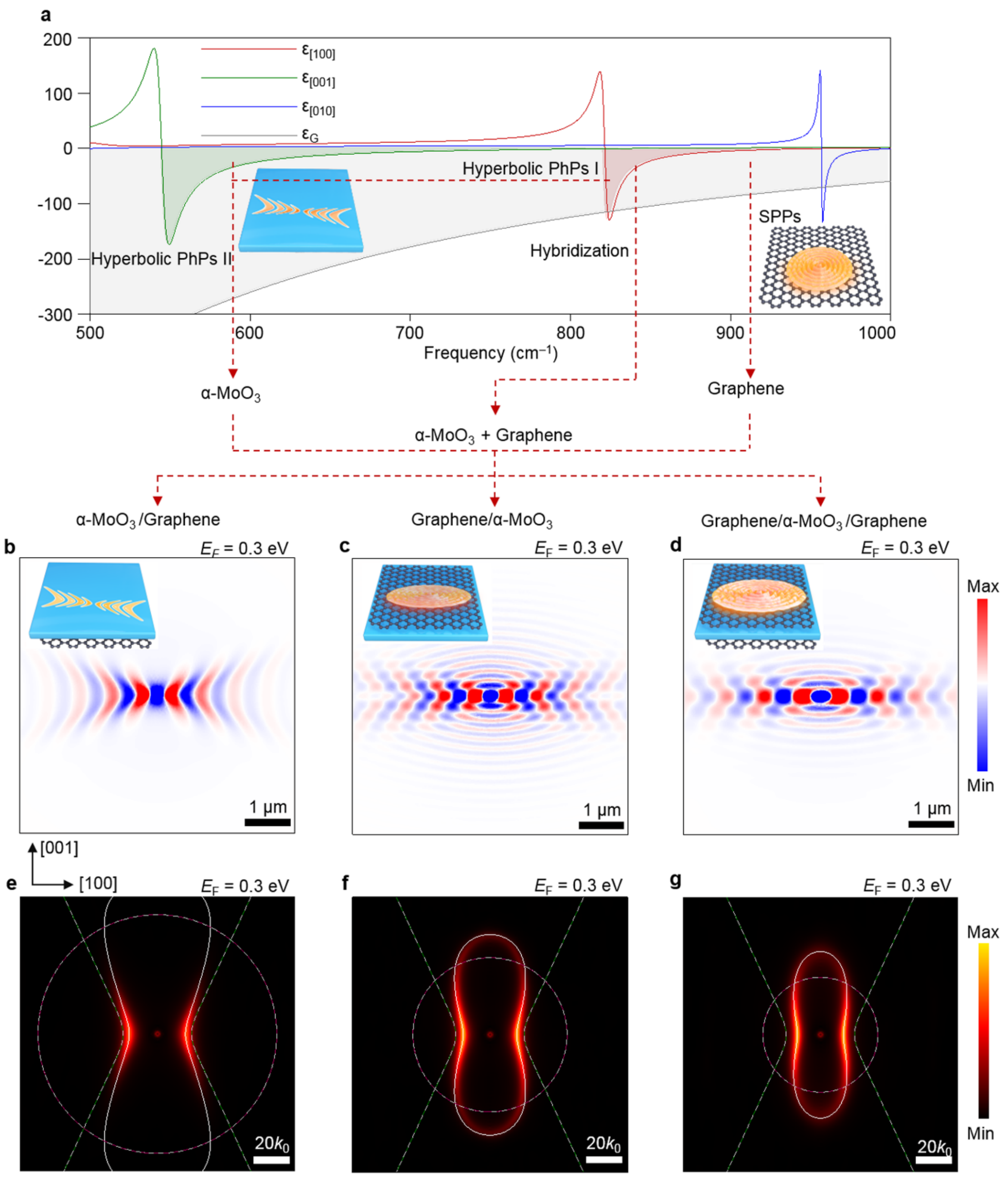

**Figure 2.** Stack-dependent topological transitions of polaritons in α-$MoO_3$ and graphene heterostructures via interface engineering. (a) Real parts of the permittivity of α-$MoO_3$ along the three principal axes and graphene. The green and red region denote the Reststrahlen band I and II of α-$MoO_3$, respectively. The light grey region represents the band where the SPPs can be excited in graphene. (b)-(d) Simulated field distributions Re($E_z$) of hybrid plasmon-phonon polaritons propagating in the α-$MoO_3$/graphene, graphene/α-$MoO_3$, graphene/α-$MoO_3$/graphene heterostructure, and (e)-(g) corresponding dispersions (fast Fourier transform (FFT) and analytical isofrequency curve) at a frequency of 905 $cm^{-1}$, respectively. The white solid curves denote the calculated dispersion bands of heterostructures in low *k*. The green and red dashed curves correspond to the dispersion curves of the graphene with different stacks and α-$MoO_3$ on the substrate, respectively. The thickness of α-$MoO_3$ is 100 nm.

We now experimentally verify the TT of hybrid plasmon-phonon polaritons. The G/$MoO_3$ heterostructures are fabricated on the $SiO_2$/Si substrate. A scattering-type near-field optical microscope (s-SNOM) is employed to map the polaritons, as schematically illustrated in Figure 3a. The optical image of the sample was shown in Figure 3b, which was prepared by selectively removing graphene (by using oxygen plasma) with an unintentional doping $E_F \approx 0.2$ eV on top of α-$MoO_3$ flake. Figure 3c depicts the near-field amplitude images measured experimentally at a frequency of 888 $cm^{-1}$. Only horizontal fringes are observed in the region of α-$MoO_3$[19], suggesting in-plane hyperbolic polaritons propagating along the [100] direction. In large contrast, in the region of G/$MoO_3$, fringes can be observed in both horizontal and vertical directions, which matches well with the simulated field distributions in Figure 3d. Similar phenomena are observed at 900 $cm^{-1}$ but with smaller fringe spacing in Figure 3e and f. Remarkably, hybrid polaritons in G/$MoO_3$ heterostructure can propagate along both [100] and [001] directions with different fringe spacing, a direct signature that the dispersion of polaritons has transited from open hyperbola to closed ellipse. The isofrequency curves of elliptical hybrid polaritons in Figure 3g show direction-dependent in-plane wavevectors, corresponding to different fringe spacing in Figure 3c and e. The in-plane wavevector increases with the angle between the edge and the [001] direction, resulting in reduced polariton wavelengths along the [001] direction.

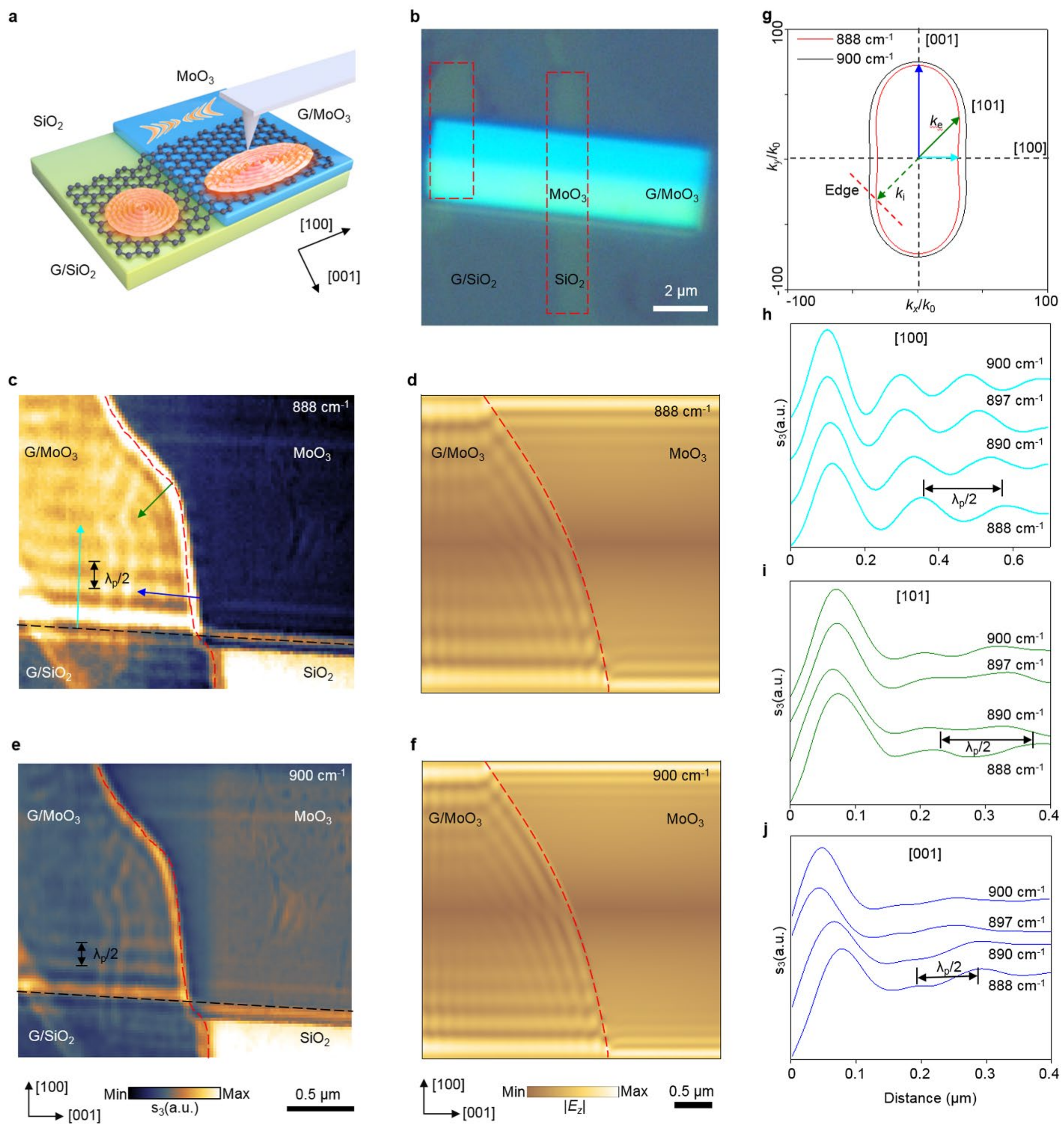

**Figure 3.** Real-space imaging of topological transition in graphene/α-$MoO_3$ heterostructure. (a) Schematic of the s-SNOM experimental scheme. (b) Optical image of graphene/α-$MoO_3$ heterostructure. Experimentally near-field amplitude s(ω) (c), (e) and simulated field (d), (f) distributions at a frequency of 888 $cm^{-1}$ and 900 $cm^{-1}$, respectively. (g) Isofrequency contours of hybrid polaritons in graphene/α-$MoO_3$ at 888 $cm^{-1}$ (black curve) and 900 $cm^{-1}$ (purple curve). (h)-(j) Line plots of measured hybrid polaritons along the cyan line, green line and blue

line in Figure 3c at different frequencies. The thickness of α-$MoO_3$ is 25 nm, and the Fermi level of graphene is 0.2 eV.

To quantitatively study the polaritonic feature, the line profiles of experimentally measured field distributions in Figure 3c at different frequencies are plotted in Figure 3h-j. The oscillating curves along the [100], [101] and [001] directions reveal the different wavevectors of propagating hybrid polaritons along these three directions, verifying the transformation of hyperbolic polaritons in G/$MoO_3$. Specifically, the polariton wavelengths $\lambda_p$ along the [100], [101], [001] directions at the frequency of 888 $cm^{-1}$ are 0.44 μm, 0.35 μm and 0.2 μm, respectively, larger than that of PhPs in α-$MoO_3$ (≈ 0.15 μm)[16]. Therefore, the wavelengths and wavevectors of hybrid polaritons can be easily tuned along different directions over a wide spectral range by just slightly changing the excitation frequency.

Graphene SPPs can be tuned by electrical gating and doping[8-10]. The dispersions and wavevectors of PhPs in α-$MoO_3$ are also sensitive to intercalated ions[18, 40, 41], and thickness-dependent[18]. We next investigate the tunability of TTs in G/$MoO_3$ heterostructure. Figure 4a illustrates the map of tailored TTs of G/$MoO_3$ heterostructure with respect to the Fermi level in graphene and the thickness of α-$MoO_3$ at 905 $cm^{-1}$. The white channel represents the Fermi levels that undergo TT for α-$MoO_3$ slabs with different thicknesses. Under this channel, hybrid polaritons can only propagate along certain angle ranges and have hyperbolic wavefront. Above this channel, hybrid polaritons can propagate in all directions even in the forbidden directions of PhPs in α-$MoO_3$, exhibiting an elliptic dispersion. Polaritons with high *k* have small wavelength, which is hard to be observed. Through this channel we can see obvious TTs from elliptical to hyperbolic dispersion regimes. With a fixed thickness of α-$MoO_3$ with 100 nm, as the wavevector of graphene plasmons decreases (i.e., the Fermi level increases), the isofrequency curve of hybrid polaritons in low $k$ changes from hyperbola to ellipse (Figure 4b). Specifically, as the Fermi level increases, the simulated fields and corresponding FFT images first exhibit hyperbolic wavefront (Figure 4c and 4f), and accompanied by the horizontal fringes emerge (Figure 4d), and finally form an elliptical wavefront (Figure 4e) with a closed dispersion contour (Figure 4h). The complete dispersion analysis and simulated field distributions can be found in Supplementary Figure S3.

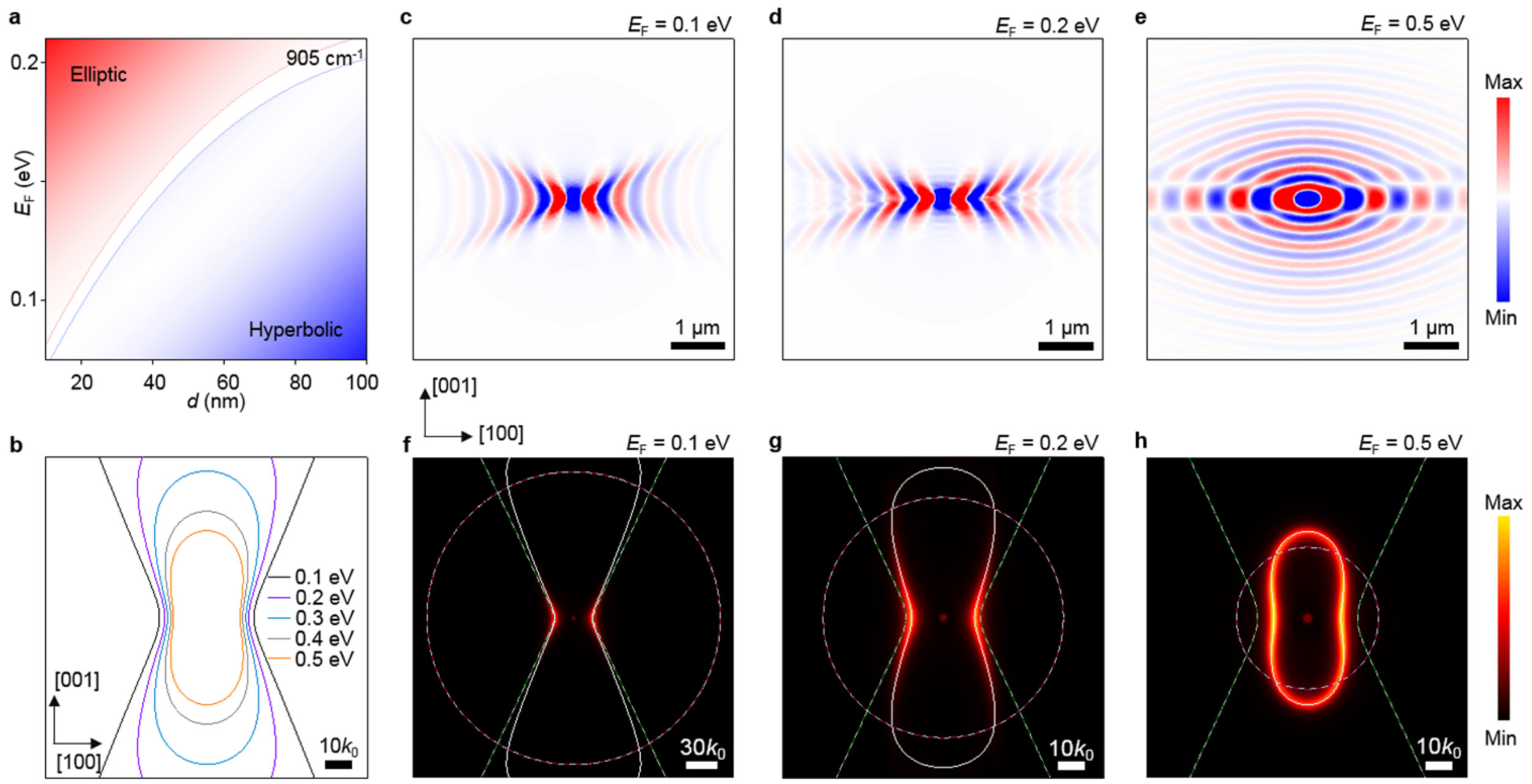

**Figure 4.** Tunable hybrid plasmon-phonon polaritons in graphene/α-$MoO_3$ heterostructures. (a) The map of the topological transition of graphene/α-$MoO_3$ heterostructure with respect to the Fermi level in graphene and the thickness of α-$MoO_3$ at a frequency 905 $cm^{-1}$. (b) Isofrequency curves for the hybrid polaritons in the graphene/α-$MoO_3$ heterostructure with different Fermi levels in graphene at a frequency 905 $cm^{-1}$. The heterostructure is stacked on a $SiO_2$/Si substrate. (c)-(e) Simulated field distributions Re($E_z$) and (f)-(h) corresponding FFT images of hybrid plasmon-phonon polaritons in graphene/α-$MoO_3$ heterostructures. The white solid curves denote the calculated dispersion bands of heterostructures in low $k$. The green and red dashed curves correspond to the dispersion curves of the graphene with different Fermi levels and α-$MoO_3$ on the substrate, respectively. The thickness of α-$MoO_3$ is 100 nm.

In addition, from Figure S3-S5, we can find that all three heterostructures can achieve TT by tuning the Fermi level of graphene with different thresholds. The hyperbolic-to-elliptical TT can be clearly observed with the increase of Fermi level to 0.5 eV and 0.2 eV in the $MoO_3$/G and G/$MoO_3$/G heterostructures, respectively. Since the modes in two graphene separated by a dielectric spacer yield smaller in-plane wavevectors than that in one layer graphene, as shown by the red dashed curves in Figure 2e-g, it is easier to achieve TT in G/$MoO_3$/G heterostructures after the coupling between SPPs and PhPs than the other two heterostructures when increasing the Fermi level. From the perspective of SPPs excited in graphene, the momentum of graphene SPPs increases as the permittivity of the environmental medium increases; Therefore, for the same Fermi level, the SPP momentum in $MoO_3$/G heterostructure is larger than that in G/$MoO_3$ heterostructure. At the TT point of G/$MoO_3$ heterostructure, graphene SPPs with circular isofrequency curve and PhPs excited in α-$MoO_3$ with hyperbolic isofrequency curve are coupled to form a flat band. However, due to the presence of the substrate, the $MoO_3$/G heterostructure requires a higher Fermi level and a lower SPP momentum to induce a TT. The results show that the stack and Fermi level can synergistically tune the TTs of such hybrid polariton systems.

The thickness of biaxial crystals is an important factor for tuning the dispersion of hyperbolic polaritons, which will thus affect the hybrid plasmon-phonon polaritons in G/$MoO_3$ heterostructures. With a fixed Fermi level, a hyperbolic-to-elliptic TT occurs as the decrease of thickness of α-$MoO_3$ (see Figure 4a). The isofrequency curve of the G/$MoO_3$ heterostructure changes from hyperbolic (black curve) to elliptical (orange curve) while the thickness decreases with the fixed Fermi level $E_F$ = 0.16 eV at a frequency 905 $cm^{-1}$ as presented in Figure 5a. For the thicknesses of 25 nm and 45 nm, the hybrid polaritons propagate in all directions with a direction-dependent wavelength, as depicted in Figure 5c and d. The corresponding FFT images (Figure 5f and g) exhibit weaker ellipticity owing to the strong anisotropy and polariton damping[27]. Since the thickness of α-$MoO_3$ decreases and the in-plane wavevector of hyperbolic PhPs in α-$MoO_3$ increases[16], a closed isofrequency contour of the hybrid polaritons can be generated gradually due to the hybridization with graphene SPPs. Inversely, as the thickness of α-$MoO_3$ increases, the decreasing in-plane wavevector of hyperbolic PhPs will gradually result in hyperbolic characteristics of the hybrid polaritons. Hybrid polaritons propagating along the [001] direction are forbidden in the heterostructure consisting of 65 nm-thick α-$MoO_3$ (Figure 5e). The corresponding FFT and the analytic dispersion in Figure 5h exhibit evident hyperbolic shapes.

To corroborate our theoretical results, we fabricate a G/$MoO_3$ heterostructure by stacking graphene on top of an α-$MoO_3$ slab with a thickness step of 65 nm and 25 nm. The measured field distributions at 888 $cm^{-1}$ and 900 $cm^{-1}$ are shown in Figure 5i and j, respectively. In the top-left region of G/$MoO_3$ heterostructure with $d$ = 25 nm, the point defect (green dashed region) and line defect (green dashed curve) respectively generates elliptical fringe patterns and oblique fringes parallel to the defects, verifying the elliptical dispersion of G/$MoO_3$ heterostructure. In the bottom-left region of G/$MoO_3$ heterostructure with $d$ = 65 nm, the polariton wavelength along the [100] direction is larger than that with $d$ = 25 nm, indicating a smaller wavevector, which is consistent with the analysis in Figure 5a. In addition, the polaritons propagating decay quickly (white dashed box in Figure 5i and j) due to the large wavevector along the [001] directions. In the right region, there only exist horizontal fringes, belonging to the hyperbolic PhPs of α-$MoO_3$. The dispersion relations for the polaritons in four different structures are plotted in Figure 5b. By carefully fitting these curves, we obtain a Femi level $E_F$ = 0.2 eV for graphene. The wavevectors of hybrid polaritons increase with decreasing thickness, indicating strong field confinement in a thinner α-$MoO_3$ flake and the thickness tunability of anisotropic polaritons.

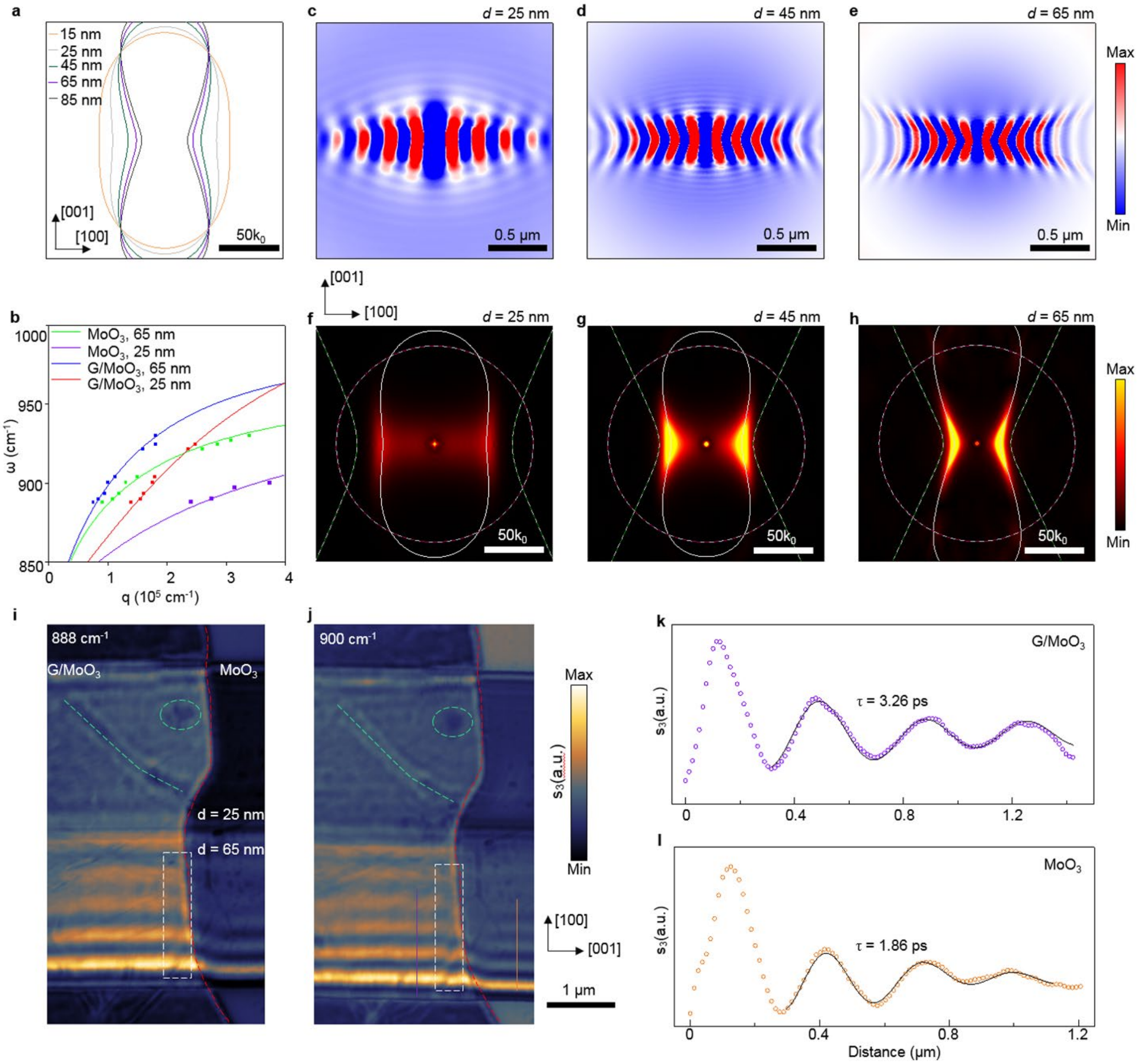

**Figure 5.** Thickness-dependent hybrid plasmon-phonon polaritons in graphene/α-$MoO_3$ heterostructures. (a) Isofrequency curves for the hybrid polaritons in the graphene/α-$MoO_3$ heterostructure with different thicknesses of α-$MoO_3$ at a frequency 905 $cm^{-1}$. The heterostructure is stacked on a $SiO_2$/Si substrate. (b) Dispersion relations for PhPs and hybrid polaritons in different structures. The solid curves represent the analytic results and the dots indicate experimental data extracted from s-SNOM. The Fermi level of graphene is 0.2 eV. (c)-(e) Simulated field distributions Re($E_z$) and (f)-(h) corresponding FFT images for the graphene/α-$MoO_3$ heterostructure. The white solid curves denote the calculated dispersion bands of heterostructures. The green and red dashed curves correspond to the dispersion curves of the graphene and α-$MoO_3$ with different thicknesses

on the substrate, respectively. Experimentally measured field distributions at (i) 888 $cm^{-1}$ and (j) 900 $cm^{-1}$. The Fermi level of graphene is 0.2 eV. Line plots of measured hybrid polaritons along the (k) purple line and (l) orange line in Figure 5j. The dots are the measured data and the solid curves are the fitting.

The interlayer coupling between graphene and α-$MoO_3$ can also significantly increase the lifetime of hybrid plasmon-phonon polaritons without complex structural fabrication. Figure 5k and l plot the line scans of measured polariton fields along the purple line and orange line in Figure 5j, respectively. Polariton lifetimes can be calculated by using $\tau = L/v_g$[19], where $L$ is the propagation length obtained from the fitting curves in Figure 5k and l, and $v_g$ is the group velocity derived from the dispersion relations in Figure 5b. Through calculations, we found that the propagation length of hybrid polaritons in G/$MoO_3$ heterostructure is longer than that in α-$MoO_3$. Although the group velocity $v_g = \partial\omega/\partial q$ of hybrid polaritons is larger than that of PhPs in α-$MoO_3$, the polariton lifetime in G/$MoO_3$ heterostructure is calculated to be $\tau = 3.26$ ps, which is 1.75 times higher than that of PhPs ($\tau = 1.86$ ps) in α-$MoO_3$ and 2.17 times higher than that of hybrid plasmon–phonon polaritons in graphene nanoribbons/h-BN heterostructure[42], revealing the ultra-low-loss character of hybrid polaritons in G/$MoO_3$ heterostructure.

## CONCLUSION

In conclusion, we have demonstrated the momentum-directed polaritons in biaxial crystals. More specifically, we theoretically and experimentally demonstrated the tailoring of hybrid plasmon-phonon polaritons, especially TTs of hybrid polaritons, by engineering the interface of graphene and α-$MoO_3$ heterostructures. Systematic calculations show that the transitions can be switched by varying the Fermi level in graphene. As the Fermi level of graphene increases, polaritons undergo a TT from open (hyperbolic) to closed (elliptical) dispersions. Furthermore, we find that the thickness of α-$MoO_3$ is another dimension of regulation, and the TT can be readily achieved with a thin $MoO_3$ slab in the presence of a lower graphene Fermi level. We have established a phase diagram of TTs of hybrid plasmon-phonon polaritons based on the Fermi level of graphene and the thickness of α-$MoO_3$, which lays a foundation for the construction of electronically-tunable polaritonic devices, optical signal processing, or neuromorphic photonic circuits based on low-loss polaritons.

## ASSOCIATED CONTENT

**Supporting Information**.

The Supporting Information is available free of charge at

Derivation of the dispersion relation of hybrid polaritons in interface engineering. Numerical simulations and experimental method. Schematic of equivalent model of interface engineering of biaxial crystals. Simulated and experimental method. Hybrid polaritons in the three suspended heterostructures. Dynamically tunable hybrid plasmon-phonon polaritons in G/$MoO_3$, $MoO_3$/G and G/$MoO_3$/G heterostructures. (PDF)

## AUTHOR INFORMATION

**Corresponding Authors**

**Qiaoliang Bao** – Department of Materials Science and Engineering, and ARC Centre of Excellence in Future Low-Energy Electronics Technologies (FLEET), Monash University, Clayton, Victoria, Australia; Email: qiaoliang.bao@gmail.com

**Huanyang Chen** – Department of Physics, Xiamen University, Xiamen, 361005, P. R. China; Email: kenyon@xmu.edu.cn

**Zhigao Dai** – Engineering Research Center of Nano-Geomaterials of Ministry of Education, Faculty of Materials Science and Chemistry, China University of Geosciences, Wuhan, 430074, P. R. China; Email: daizhigao@cug.edu.cn

**Authors**

**Yali Zeng** – Department of Physics, Xiamen University, Xiamen, 361005, P. R. China

**Qingdong Ou** – Department of Materials Science and Engineering, and ARC Centre of Excellence in Future Low-Energy Electronics Technologies (FLEET), Monash University, Clayton, Victoria, Australia

**Lu Liu** – Engineering Research Center of Nano-Geomaterials of Ministry of Education, Faculty of Materials Science and Chemistry, China University of Geosciences, Wuhan, 430074, P. R. China

**Chunqi Zheng** – Department of Electrical and Computer Engineering, National University of Singapore, Singapore, 117583, Singapore

**Ziyu Wang** – Institute of Materials Research and Engineering Agency for Science Technology and Research (A*STAR), Singapore, 138634, Singapore

**Youning Gong** – Institute of Microscale Optoelectronics, Shenzhen University, Shenzhen, 518060, P. R. China

**Xiang Liang** – School of Energy and Power Engineering, Wuhan University of Technology, Wuhan, 430063, P. R. China

**Yupeng Zhang** – Institute of Microscale Optoelectronics, Shenzhen University, Shenzhen, 518060, P. R. China

**Guangwei Hu** – Department of Electrical and Computer Engineering, National University of Singapore, Singapore, 117583, Singapore

**Zhilin Yang** – Department of Physics, Xiamen University, Xiamen, 361005, P. R. China

**Cheng-Wei Qiu** – Department of Electrical and Computer Engineering, National University of Singapore, Singapore, 117583, Singapore

**Present Address**

**Yali Zeng** – Department of Electrical and Computer Engineering, National University of Singapore, Singapore, 117583, Singapore

**Author Contributions**

† Y.Z. and Q.O. contributed equally to this work.

**Author Contributions**

Q.B. and H.C. conceived the original concept with kind discussion with Y.Z., Q.O. and Z.D.. Q.B. and H.C. supervised the project. Z.D. carried out the near-field imaging experiments with the help of Q.O. and Z.W.. Y.Z. performed the modelling and data analysis with the supervision of H.C. Y.Z. carried out the simulations with the help of L.L.. Z.W. , Q.O. and Z.D. contributed to the material synthesis and sample fabrication. Y.Z., Q.O., L.L., C.Z., G.H., L.L., C.Q., Q.B., H.C.and Z.D. participated in data analysis and co-wrote the manuscript. All authors edited the paper.

**Notes**

The authors declare no competing financial interest.

**ACKNOWLEDGMENT**

We thank Weiliang Ma for discussions and help with the sample fabrication. We acknowledge the support from the National Natural Science Foundation of China (No. 52172162, 11874311 and 92050102), the National Key Research & Development Program (No. 2016YFA0201900 and 2020YFA0710100), the Australian Research Council (ARC, CE170100039 and DE220100154), and the China Scholarship Council (No. 202006310049). C.-W. Q. acknowledges financial support from A*STAR Pharos Program (grant number 15270 00014, with project number R-263-000-B91-305). Z. D. acknowledges support from the Fundamental Research Funds for the Central

Universities, China University of Geosciences (Wuhan) (No. 162301202610). H.C. acknowledges support from the Fundamental Research Funds for the Central Universities (Grant No. 20720200074). This work was performed in part at the Melbourne Centre for Nanofabrication (MCN) in the Victorian Node of the Australian National Fabrication Facility (ANFF).